\documentstyle[prl,aps,psfig,multicol]{revtex}
\begin{document}
\draft
\begin{multicols}{2}
\title{Reply to the Comment on ''Electric Field Scaling at a $B=0$ 
Metal-Insulator Transition in Two Dimensions''}
\maketitle
In their Comment \cite{ismail} on our recent Letter \cite{efield} on 
electric field scaling at the $B=0$ conductor-insulator transition 
in silicon MOSFETs, Ismail and coauthors report experimental data that they 
attribute 
to the occurence of a
transition to a conducting phase in another silicon-based material, a Si/SiGe 
heterostructure.

If correct, it is indeed a very interesting finding which would establish 
the occurence of this unexpected transition in a material other 
than high-mobility silicon MOSFETs.  We point out, however, that the claim is based 
on little data.  Measurements were taken at only three temperatures (0.4, 
1.2 and 4.2 K) for 
each electron density, and a single crossing 
point is not clearly demonstrated in  Fig.1(a).  In fact, there is not 
even enough 
data to preclude a possible 
maximum in conductivity between 0.4 and 4.2 K.  
It is important to note also that the electron densities range from $0.55 n_c$ 
to $3.3 n_c$, so that the scaling curves shown in Fig. 1 (b) encompass values of 
$\delta \equiv 
|n-n_c|/n_c$ from 0.15 to 2.3, well 
outside the critical region where one would 
expect scaling to be obeyed.

Ismail {\it et al.} claim that the conductivity at the critical point is not 
universal: the conductivity of their devices at the critical point is 
$100 e^2/h$, two orders of magnitude larger than in silicon.  We note 
that MOSFETs which differ in mobility by a factor 
of nearly 4 (20,000 to 75,000 cm$^{2}/$Vs) were found to have quite similar critical 
conductivities near $e^2/3h$ (at about 0.1 K), implying a universal value, at least for electrons 
in silicon inversion layers on Si-SiO$_2$ interfaces\cite{efield,krav94,krav95}.

A conductivity that increases with decreasing temperature does not by itself 
establish the existence of a conducting phase.  An example is provided by 
low-mobility silicon MOSFETs, where an increase in the conductivity 
in the restricted range from 5 K to 1.2 K was 
attributed to temperature-dependent Coulomb scattering in the {\it insulating} 
phase \cite{wheeler}.  Although there may well prove to be a conductor-insulator 
transition in Si/SiGe heterostructures, additional data are needed 
to establish the validity of this claim.
\vskip 12pt
\noindent S. V. Kravchenko, D. Simonian, and M. P. Sarachik

Physics Department

City College of the City University of New York

New York, New York 10031

\end{multicols}
\end{document}